\documentclass[aps,prl,twocolumn,showcaps,superscriptaddress,showpacs,amsmath,amssymb]{revtex4}

\usepackage{graphicx}
\usepackage{dcolumn}
\usepackage{bm}
\usepackage[dvips]{epsfig}
\usepackage{color}
\usepackage{upgreek}
\usepackage{color}

\newcommand{\unit}[1]{\;\mathrm{#1}}

\newcommand{\kT}{\ensuremath{k_{\rm{B}}T}}

\newcommand{\ps}{\ensuremath{p^{\rm s}}}

\newcommand{\js}{\ensuremath{j^{\rm s}}}

\newcommand{\nus}{\ensuremath{\nu^{\rm s}}}
\newcommand{\nucg}{\ensuremath{\tilde{\nu}^{\rm s}}}
\newcommand{\stot}{\ensuremath{\Delta s_{\rm tot}}}
\newcommand{\sred}{\ensuremath{\Delta\tilde{s}_{\rm tot}}}
\newcommand{\nprime}{\ensuremath{\tilde{n}}}
\newcommand{\xcg}{\tilde{\vec{x}}}

\newcommand{\LandauO}[1]{\ensuremath{\mathcal{O}\!\left(#1\right)}}

\renewcommand{\vec}[1]{\mathbf #1}

\begin{document}

\title{Role of Hidden Slow Degrees of Freedom in the Fluctuation Theorem}

\author{J. Mehl}
\affiliation{2. Physikalisches Institut, Universit\"at Stuttgart, Pfaffenwaldring 57, 70569 
Stuttgart, Germany}
\author{B. Lander}
\affiliation{II. Institut f\"ur Theoretische Physik, Universit\"at Stuttgart, Pfaffenwaldring 57, 
70550 Stuttgart, Germany}
\author{C. Bechinger}
\affiliation{2. Physikalisches Institut, Universit\"at Stuttgart, Pfaffenwaldring 57, 70569 
Stuttgart, Germany}
\affiliation{Max-Planck-Institute for Intelligent Systems, Heisenbergstrasse 3, 70569 Stuttgart, 
Germany}
\author{V. Blickle}
\affiliation{2. Physikalisches Institut, Universit\"at Stuttgart, Pfaffenwaldring 57, 70569 
Stuttgart, Germany}
\affiliation{Max-Planck-Institute for Intelligent Systems, Heisenbergstrasse 3, 70569 Stuttgart, 
Germany}
\author{U. Seifert}
\affiliation{II. Institut f\"ur Theoretische Physik, Universit\"at Stuttgart, Pfaffenwaldring 57, 
70550 Stuttgart, Germany}


\begin{abstract}
The validity of the fluctuation theorem for entropy production as deduced from the observation
of trajectories implicitly requires that all slow degrees of freedom are accessible. We 
experimentally investigate the role of hidden slow degrees of freedom in a system of two 
magnetically coupled driven colloidal particles. The apparent entropy production based on the 
observation of just one particle obeys a fluctuation theorem-like symmetry with a slope of 1 in 
the short time limit. For longer times, we find a constant slope, but different from 1. We present 
theoretical arguments for a generic linear behavior both for small and large apparent entropy 
production but not necessarily throughout. By fine-tuning experimental parameters, such an 
intermediate nonlinear behavior can indeed be recovered in our system as well.
\end{abstract}


\maketitle


{\it Introduction.}---Basic concepts of thermodynamics and statistical physics implicitly rest on  
a separation of all degrees of freedom into observable and non-observable  ones. Heat exchange, 
for example, is associated with the myriads of fast degrees of  freedom which are not resolved 
dynamically, whereas work typically involves  a few controlled, slow degrees of freedom. In 
systems without a clear-cut  time-scale separation, ambiguities and inconsistencies may arise if 
such concepts are still applied naively. Here we explore this issue for one of the  arguably most 
relevant concepts, entropy production $\stot$, for non-equilibrium steady states (NESS). For 
such states, the fluctuation theorem (FT) refers to a remarkable symmetry that quantifies the 
probability $p$ of observing trajectories with negative total entropy production as 
\begin{equation}
 \label{eq:FT}
 \ln\left[p(\stot)/p(-\stot)\right] = \alpha\stot
\end{equation}
with $\alpha = 1$ and Boltzmann's constant set to unity~\cite{evans93,galla95,kur98,lebo99,sei05}.

The FT has been proven for two types of dynamics. First, for deterministic dynamics the proof
rests on the {\it chaotic hypothesis}, including time reversibility and a phase-space contraction
associated with dissipation~\cite{galla95}. Second, in stochastic dynamics the FT requires the 
concept of entropy production along trajectories and can be proven for Markovian 
systems~\cite{kur98, lebo99,sei05}. The latter dynamics applies to experiments on driven colloidal 
particles~\cite{wang02, spe07} and a harmonic oscillator coupled to a thermal bath~\cite{dou06}.
Experimental tests of FT-like symmetries have also been reported for Rayleigh-B\'{e}nard 
convection~\cite{cili98}, turbulent flow~\cite{cili04}, granular matter~\cite{feit04}, and  
self-propelled particles~\cite{kum11}. For these systems the appropriate class of dynamics
is less obvious and hence the status regarding the assumptions of the FT is unclear {\it a priori}.
One should also appreciate that the measured observable for some of these systems is typically 
not $\stot$ directly, but rather some dimensionful quantity, like, e.g., the injected or 
dissipated work~\cite{feit04, auma01}, which requires a temperature for a unique conversion to 
entropy. Strictly speaking, the FT is thus verified only if this temperature can be determined 
independently and if it leads to $\alpha = 1$. Complementary, in more recent reports, the validity 
of the FT (with $\alpha = 1$) is assumed and used to gain information about such a dimensionful 
factor connecting the actual observable with entropy production~\cite{haya10, suzu11}.

For stochastic dynamics, the proof of the FT with $\alpha = 1$ rests on a time-scale separation.
Fast degrees of freedom contribute to an effectively white noise leading then to a Markovian
dynamics of the slow degrees of freedom. Entropy production can be deduced from observing the
dynamics of {\it all} slow degrees of freedom. If some of these degrees of freedom are not, or
cannot, be observed the inferred entropy production is only an apparent one for which the status of
an FT-like symmetry is unclear {\it a priori}. Theoretical efforts to describe coarse-graining in 
general have been restricted so far to the case of well separated time 
scales~\cite{raha07,pugli10,espo12}, and how such coarse-graining affects bounds on dissipated 
work~\cite{kawa07}. In the framework of electronic devices FT-like symmetries for currents have 
been discussed in Refs.~\cite{utsu10,gane11,cue11}.

In the present paper, we investigate the role of hidden slow degrees of freedom on apparent
entropy production for a paradigmatic system with two magnetically coupled driven colloidal
particles.


{\it Apparent Entropy Production.}---The total entropy production $\stot$ is given by the sum of 
the entropy changes of the heat bath and system~\cite{sei05}
\begin{equation}
 \label{eq:s_tot1}
 \stot = Q/T+\ln\left[\ps(\vec{x}^0)/\ps(\vec{x}^{t})\right],
\end{equation}
where $Q=\int_0^t d\tau\sum_{i=1}^n\dot{x}_i(\tau)F_i(\vec{x}(\tau))$ is the heat transfer of all
$n$ degrees of  freedom $\vec{x}\equiv(x_1\ldots x_n)$ to the solvent at temperature $T$. Here,
$\dot{x}_i$ is the actual velocity and $F_i$  is the total force  acting on the $i^{\rm th}$ degree
of freedom. The change of the system's entropy includes the stationary probability distribution 
$\ps(\vec{x}^0)$ ($\ps(\vec{x}^t)$) of finding the initial (final) state of the system along the 
trajectory of length $t$. In a NESS, the system satisfies the stationary Smoluchowski equation $0 
=  -\sum_{i}\partial_{x_i}\js_i(\vec{x})$, where the probability current $\js_i(\vec{x}) = 
\ps(\vec{x})\nus_i(\vec{x})$ is given as the product of $\ps$ and the mean local  
velocity~\cite{sei05}
\begin{equation}
 \label{eq:nus}
 \nus_i(\vec{x}) \equiv D_0\left[F_i(\vec{x})/T-\partial_{x_i}\ln{\ps(\vec{x})}\right]
\end{equation}
with $D_0$ the bare diffusivity. Multiplying Eq.~\eqref{eq:nus} with $\dot{x}_i$ and integrating
over time yields the total entropy production as given in Eq.~\eqref{eq:s_tot1},
\begin{equation}
 \label{eq:s_tot2}
 \stot=\int\limits_{0}^{t}\,d\tau\sum\limits_{i=1}^{n}\dot{x}_i(\tau)\nus_i(\vec{x}(\tau))/D_0,
\end{equation}
where the sum involves all $n$ degrees of freedom $\vec{x}$. If only the first $\nprime$ of these,
$\tilde{\vec{x}} \equiv (x_1\ldots x_{\nprime})$, are accessible, an observer is forced to deduce
all information from these trajectories. The actual velocities $\dot{\tilde{\vec{x}}}$ can still be
measured correctly, whereas the mean local velocity obtained from the accessible trajectories is
\begin{equation}
 \nucg_i(\tilde{\vec{x}}) \equiv
\int\nus_i(\vec{x})\ps(\hat{\vec{x}}|\tilde{\vec{x}})d\hat{\vec{x}}
\end{equation}
with the conditional probability $\ps(\hat{\vec{x}}|\tilde{\vec{x}})$ for $\hat{\vec{x}} \equiv
(x_{\nprime+1},\ldots, x_n)$ at fixed $\tilde{\vec{x}}$. Hence, the apparent entropy production
becomes
\begin{equation}
 \label{eq:stot_red}
 \sred =
\int\limits_{0}^{t}\,d\tau\sum\limits_{i=1}^{\nprime}\dot{x}
_i(\tau)\nucg_i(\xcg(\tau))/D_0,
\end{equation}
where the sum runs over the $\nprime$ accessible degrees of freedom only. In this Letter, we
investigate the conditions under which this quantity obeys a FT-like symmetry.


{\it Experiment.}---We have created two non-overlapping toroidal traps with radius $R =
3.5\unit{\upmu m}$ and a center-center distance of $17\unit{\upmu m}$ by a single laser beam
$(\lambda = 1070\unit{nm})$ which was deflected on a galvanometric mirror unit (for details refer 
to~\cite{fau95,mehl10}). Each trap contained a single paramagnetic colloidal particle with a
$2.6\unit{\upmu  m}$ radius (Microparticles, Berlin) and labeled by an index $i = 1,2$ (see
Fig.~1(a)). The traps are approximately $50\unit{\upmu m}$ away from the lower surface; therefore
hydrodynamic interactions with the walls are negligible~\cite{lutz06}. The scanning frequency was
adjusted to $41\unit{Hz}$, which leads to quasistatic tangential forces $f_i$ acting on the $i^{\rm
th}$ particle along the toroidal traps whose amplitude depends on the laser intensity. In our
experiments, the time for a full revolution of each particle was adjusted to $10\unit{s}$.
Synchronized with the scanning motion of the laser beam, its intensity was sinusoidally modulated
with an acousto-optic modulator which finally leads to an effective optical potential $U_i(x_i) =
V_i(x_i)-f_ix_i$ with $V_i(x_i) = V^0_i\sin(x_i)$, where $x_i$ is the particle position along the
trap circumference in units of $\pi R$ (see Fig.~1(a)). Accordingly, both particles reach NESS,
where $f_i$ and $V^0_i$ can be controlled independently.

\begin{figure}
 \includegraphics[width=0.85\linewidth]{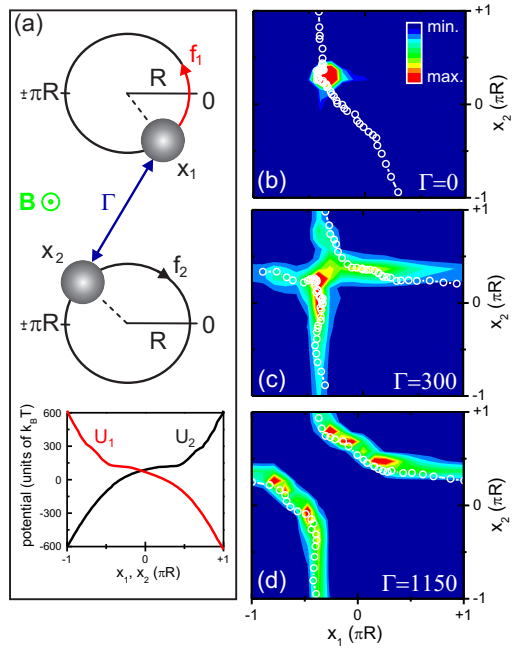}
 \caption{(color online) (a)~Schematic representation of the system and measured tilted potentials
 ($V_1^0 = V_2^0 = 181\unit{\kT}$ and  $f_1 = -f_2 = 57\unit{\kT/\upmu m}$ calculated  
 via~\cite{bli07}). (b)--(d)~Stationary probability  distribution $\ps(x_1,x_2)$ for different  
 plasma parameters $\Gamma$. The white circles indicate typical trajectories.   \label{fig1}}
\end{figure}

A coupling between the two NESS is obtained by a static homogeneous magnetic field $B$ applied
perpendicular to the sample plane. This field induces magnetic moments $m$ to the particles leading
to a repulsive dipolar particle interaction $W(x_1,x_2) = (\mu_0/4\pi)\, m^2/r^3(x_1, x_2)$. Here,
$\mu_0$ is the magnetic constant and $r$ the  particle distance. For small magnetic fields $(B
\leqslant 40\unit{mT})$, as in our experiments, the magnetic moment scales as $m \approx (\gamma
m_0/3)B$ with $\gamma = 30\unit{T^{-1}}$ and $m_0 = 5.9 \times 10^{-13}\unit{Am^2}$~\footnote{We
determine $\gamma$ and $m_0$ by measuring the drag force when a particle is subjected to a magnetic
field gradient.}. The strength of the dipolar coupling can be conveniently characterized by a
dimensionless plasma parameter $\Gamma \equiv \Delta W/(\kT)$, where $\Delta W$ corresponds to the
difference of the coupling at the smallest and largest particle distance $r$. Additional particle
interactions, e.g., hydrodynamic coupling or optical binding, are negligible at the chosen trap
separations as confirmed by the independent motion of the particles in the absence of a magnetic
field.

{\it Results.}---We investigate the effect of coupling by preparing two identical NESS, where the 
potential minima face each other. Figures~\ref{fig1}(b)--(d) show the stationary probability
distribution $\ps(x_1,x_2)$ as color coded background and as white circles one exemplary 
trajectory. In the absence of coupling, the peak in $\ps$ corresponds to the flattest part in the 
potentials, where the particles slow down and therefore are most likely to be found (see 
Fig.~\ref{fig1}(b)). Under strong coupling conditions (see Fig.~\ref{fig1}(d)), the repulsion 
hinders the particles from coming close to each other. The peak of $\ps$ vanishes since the 
approach of one particle kicks the other one away, leading to motion like a Newton's cradle. In 
the intermediate regime (see Fig.~\ref{fig1}(c)), the coupling interaction is comparable with the 
energy loss acquired while the  particle moves along half a circle. Here the full interplay 
between drift, diffusion, and interaction has to be taken into account and we expect the influence 
of hidden degrees of freedom on the FT to be most prominent. Therefore, we concentrate on $\sred$ 
associated with the motion of only the first particle, which represents the observed degree of 
freedom, whereas the coupling allows us to uniquely control the influence of the second particle, 
which acts as a hidden degree of freedom.


\begin{figure}
 \includegraphics[width=\linewidth]{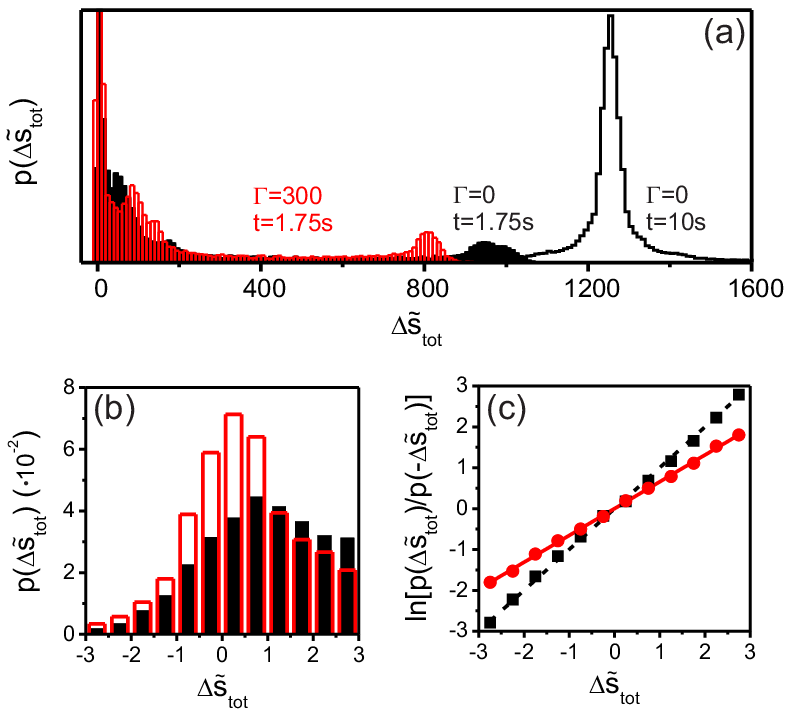}  
 \caption{(color online) (a)~Distribution of the apparent entropy production $p(\sred)$ for 
 different trajectory  lengths $t$ and plasma parameters $\Gamma$. (b)~Section of   previous 
 histograms around $\sred=0$. (c)~Corresponding $\ln\left[p(\sred)/p(-\sred)\right]$ as a function 
 of $\sred$. The dashed black line has the theoretically predicted slope $1$,  whereas the red 
 line is a linear fit with slope $\alpha = 0.65$. \label{fig2}}
\end{figure}

The black histograms (closed bars and line) in Fig.~\ref{fig2}(a) show the distribution of the 
apparent entropy production $p(\sred)$ in the absence of coupling obtained for trajectories of 
length $t=1.75\unit{s}$ and $10\unit{s}$, respectively. The  peaked distribution shifts with 
elapsing time to the right with peak height maxima occurring at positions which correspond to the 
energy loss associated with full revolutions of the particle, $2\pi Rf = 1250\unit{\kT}$. To 
investigate the FT, rare events with negative entropy production have to be sampled with high 
accuracy. This constrains the maximal trajectory length $t$ to approximately $2\unit{s}$ and the 
range within the FT can be tested to $\pm3$. Figure~\ref{fig2}(b) shows this section of the black 
(closed bars) histogram around $\sred=0$. The  excellent agreement between the logarithm of the 
probability ratio  $p(\sred)/p(-\sred)$, black squares in Fig.~\ref{fig2}(c), and the black dashed 
line with a slope of $1$ confirms the validity of Eq.~\eqref{eq:FT} for uncoupled states. The red 
histogram (open bars) in Figs.~\ref{fig2}(a) and \ref{fig2}(b) demonstrates the situation for 
coupled states. Most prominent is the enhanced probability at $\sred = 0$. Since the red dots in 
Fig.~\ref{fig2}(c) do not agree with the dashed line of slope $1$ this apparent entropy production 
does not obey the FT. Rather a linear relation as given by Eq.\eqref{eq:FT} with $\alpha \simeq 
0.65$ is found.

\begin{figure}
 \includegraphics[width=\linewidth]{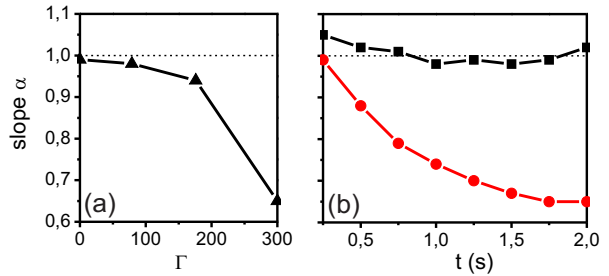}
 \caption{(color online) (a) Slope $\alpha$ vs plasma parameter $\Gamma$ for $t = 1.75\unit{s}$.
 (b) Slope $\alpha$ for  different trajectory lengths $t$. The black squares correspond to $\Gamma 
 = 0$ and the red dots to $\Gamma = 300$. The  deviation of the black squares from $\alpha = 1$ 
 (black dashed line) determines the statistical errors to be less than $5\%$. \label{fig3}}
\end{figure}

In additional experiments, we observe linear relations according to Eq.~\eqref{eq:FT} with 
different slopes $\alpha$, which depend on two parameters: (i) the plasma parameter $\Gamma$, and
(ii) the trajectory length $t$, as shown in Figs.~\ref{fig3}(a) and \ref{fig3}(b). Clearly, the FT 
is confirmed for arbitrary trajectory lengths in uncoupled states (black squares in 
Fig.~\ref{fig3}(b)). The obvious dependence of $\alpha$ on $\Gamma$ resembles the transition from 
an uncoupled to a coupled state. For $\Gamma = 300$, $\alpha$ decays with increasing length $t$, 
from $1$ to $0.65$. A similar time dependence has been also observed in~\cite{haya10}; however, 
there it was not identified as an inherent feature of hidden degrees of freedom. In additional 
measurements performed for coupling two different NESS, we also found such a linear relation. 
Therefore, we exclude symmetry as the sole origin of this behavior.

{\it Discussion.}---First, we explain why for $t \rightarrow 0$ the slope $\alpha$ approaches $1$. 
In general, deviations from the FT must be caused by the interaction with the hidden particle. In a
short time expansion to lowest order in $t$, we can neglect changes in the interaction force 
during the motion of the observed particle. Thus, the interaction force entering 
Eq.~\eqref{eq:stot_red} through $\nucg$ becomes constant in this limit. The apparent entropy 
production then becomes equivalent to that of an effective one-particle system subject to a 
Markovian dynamics with mean local velocity $\nucg$, which trivially obeys the  FT. This effective 
description is valid only to lowest order in time since taking into account 
higher order terms would include contributions arising from the correlated motion of the observed 
and the hidden particle.


Although we are able to qualitatively understand the influence of coupling it remains a surprising
feature why, in all of our experiments presented so far, only the slope of the FT is affected by 
the coupling strength while the linear  relation Eq.~\eqref{eq:FT} remains untouched. In order to 
elucidate this result we define the function
\begin{equation}
 \label{eq:function}
 f(\sred) \equiv \ln\left[p(\sred)/p(-\sred)\right],
\end{equation}
which we assume to be analytic. First, we note that $f$ is antisymmetric by construction, and thus 
for small entropy productions, $\sred \ll 1$, $f$ trivially must be linear up to corrections of 
third order or higher~\cite{pugl05}. Second, we discuss $f$ for large entropy productions, $\sred 
\gg 1$. Solving Eq.~\eqref{eq:function} for $p(-\sred)$ and integrating over all $\sred$ yields
\begin{equation} 
 \label{eq:norm}
 \int_{-\infty}^{+\infty}p(\sred)\,e^{-f(\sred)}\,d\sred = 1,
\end{equation}
by normalization. We assume that $p(\sred)$ does not decay faster than a Gaussian as we have
observed in all our measurements. For any quantity consisting of independent contributions the 
central limit theorem would imply a Gaussian. Any correlation will typically lead to an even 
slower decay. Under this assumption convergence of the integral in Eq.~\eqref{eq:norm} requires 
that $f(\sred) = \LandauO{\sred^2}$. Since, in addition, $f$ is antisymmetric we expect the 
generic asymptotic behavior to be linear, $f(\sred) \sim \sred$, with a slope generally different 
from the one for small $\sred$. 

\begin{figure}
 \includegraphics[width=\linewidth]{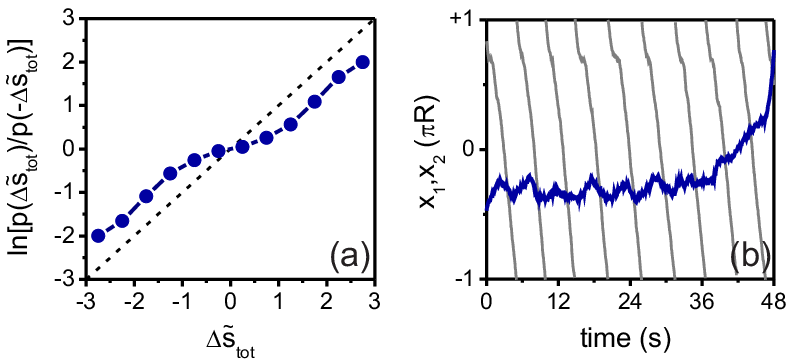}
 \caption{(color online) (a)~$\ln\left[p(\sred)/p(-\sred)\right]$ as a function of $\sred$ for  $t
 = 3\unit{s}$. (b)~Typical  trajectories of the observed (blue/thick line) and hidden particle 
 (gray/thin line). \label{fig4}}
\end{figure}

Summarizing these arguments, we expect a linear function both for small and for large entropy
production for any time $t$. For intermediate entropy production this reasoning leaves the 
possibility of a nonlinear behavior. Even though we have found a constant slope for most 
experimental parameters, by fine-tuning the system and plasma parameter, we can observe an 
obviously nonlinear result, as shown in Fig.~\ref{fig4}(a).


In contrast to the previous data, here, the two particles are subjected to quite different
potentials whereas the driving forces remain untouched. The potential of the hidden particle is
adjusted such that it circulates freely along the torus ($V^0_2 = 71\unit{\kT}$) whereas a  deep
minimum ($V^0_1 = 262\unit{\kT}$) remains in the tilted potential $U_1$ of the observed particle,
which, for $\Gamma = 0$, it is not able to leave. The latter's motion sets in only when the 
coupling helps it to surmount the potential barrier. This mode is identified in a typical 
trajectory shown in Fig.~\ref{fig4}(b).  The hidden particle (gray/thin line) moves with a period 
of $4\unit{s}$ and almost constant velocity along $U_2(x_2)$. Around $x_2 = 0.75$ it slightly 
slows down due to interaction. The reaction of the observed particle (blue/thick line) is more 
pronounced since locally it is confined within a potential minimum at $x_1 = -0.35$ and there the 
interaction forces are dominant displacing it along positive $x_1$.  The apparent oscillations in 
the trajectory originate from the fact that not every time the observed particle is pushed (by the 
hidden one) this action results in a surmounting of the potential barrier. In most of the cases, 
the particle just relaxes to its original position. We observe that nonlinearities in the 
intermediate regime of $\sred$ are most pronounced when the trajectory length matches 
approximately the oscillation period.


{\it Concluding Perspectives.}---We have investigated the influence of hidden slow degrees of
freedom on the FT. In our experiments, we typically find that a FT-like symmetry is preserved, 
however, with a different slope which depends, in particular, on the length of the observed
trajectories. Consequently, in any experiment, where hidden slow degrees of freedom cannot be ruled
out {\it a priori}, an observed linear behavior cannot be used to infer quantities by implicitly
assuming $\alpha=1$. Theoretically, we have argued that a slope of 1 is to be expected for short
trajectories while, for any length, for both small and large entropy production a constant slope 
should be generic. Classifying theoretically the conditions for finding an almost constant slope 
over the full range, as we often did in our experiments, remains a task for future work. Likewise, 
it will be important to explore, both in theory and experiments, how hidden slow degrees of 
freedom affect other quantities, like work and heat, their exact relations, and the 
fluctuation-dissipation theorem for NESS.


We thank C. Groben for characterizing the particles and D. Hartich and J.~R. Gomez-Solano for 
interesting discussions. V.~B. was supported by the Deutsche Forschungsgemeinschaft (Grant No. BL 
1067).


\end{document}